\include{graphicsx} 

\documentclass[12pt,preprint]{aastex}

\usepackage{emulateapj5}

\slugcomment{Submitted June 23, 2003; accepted August 19, 2003, to appear in the December 10, 2003 issue (Vol 599) of the ApJ}

\shorttitle{The Complex Theta 1 Ori B System}
\shortauthors{Close et al.}

\begin{document}


\title{High Resolution Images of Orbital Motion in the Trapezium Cluster: First Scientific Results from the MMT Deformable Secondary Mirror Adaptive Optics System{\footnotemark}}

\footnotetext{A portion of the results presented here made use of the of MMT Observatory, a facility jointly operated by the University of Arizona and the Smithsonian Institution.}


\author{Laird M. Close$^1$, Francois Wildi$^1$, Michael Lloyd-Hart$^1$, Guido Brusa$^{12}$, Don Fisher$^1$, Doug Miller$^1$, Armando Riccardi$^2$, Piero Salinari$^2$, Donald W. McCarthy$^1$, Roger Angel$^1$, Rich Allen$^1$, H.M. Martin$^1$, Richard G. Sosa$^1$, Manny Montoya$^1$, Matt Rademacher$^1$, Mario Rascon$^1$, Dylan Curley$^1$, Nick Siegler$^1$, Wolfgang J. Duschl$^3$ }

\email{lclose@as.arizona.edu}

\affil{$^1$Steward Observatory, University of Arizona, Tucson, AZ 85721}
\affil{$^2$Arcetri Observatory, Universita degli Studi di, Firenzi, I-50125 Italy}
\affil{$^3$Institut fur Theoreetische Astrophysik, Universitat Heidelberg, D-69121, Germany}



\begin{abstract} 

 We present the first scientific images obtained with a deformable
secondary mirror adaptive optics system. We utilized the 6.5m MMT AO
system to produce high-resolution (FWHM=0.07$\arcsec$) near infrared
(1.6$\mu m$) images of the young ($\sim 1$ Myr) Orion Trapezium
$\theta^{1}$ Ori cluster members. A combination of high spatial
resolution and high signal to noise allowed the positions of these
stars to be measured to within $\sim 0.003\arcsec$ accuracies. We also
present slightly lower resolution (FWHM$\sim$0.085$\arcsec$) images
from Gemini with the Hokupa'a AO system as well. Including previous
speckle data \citep{wei99}, we analyze a six year baseline of high-resolution
observations of this cluster. Over this baseline we are sensitive to
relative proper motions of only $\sim0.002\arcsec$/yr (4.2 km/s
at 450 pc). At such sensitivities we detect orbital motion in the very
tight $\theta^{1}$ Ori $B_{2}B_{3}$ (52 AU separation) and
$\theta^{1}$ Ori $A_{1}A_{2}$ (94 AU separation) systems. The relative
velocity in the $\theta^{1}$ Ori $B_{2}B_{3}$ system is $4.2\pm2.1$
km/s. We observe $16.5\pm5.7$ km/s of relative motion in the
$\theta^{1}$ Ori $A_{1}A_{2}$ system. These velocities are consistent with those independently observed by \cite{sch03} with speckle interferometry, giving us confidence that these very small ($\sim 0.002\arcsec$/yr) orbital motions are real. All five members of the
$\theta^{1}$ Ori $B$ system appear likely gravitationally bound
($B_{2}B_{3}$ is moving at $\sim1.4$ km/s in the plane of the sky
w.r.t. $B_{1}$ where $V_{esc} \sim 6$ km/s for the B group). The very
lowest mass member of the $\theta^{1}$ Ori $B$ system ($B_{4}$) has
$K^\prime \sim 11.66$ and an estimated mass of $\sim 0.2 M_{\sun}$. There was very little motion ($4\pm 15$ km/s) detected of $B_4$ w.r.t $B_1$ or $B_2$, hence $B_4$ is possibly part of the $\theta^{1}$ Ori $B$ group. We
suspect that if this very low mass member is physically associated
it most likely is in an unstable (non-hierarchical)
orbital position and will soon be ejected from the group. The
$\theta^{1}$ Ori $B$ system appears to be a good example of a star
formation ``mini-cluster'' which may eject the lowest mass members of
the cluster in the near future. This ``ejection'' process could play a
major role in the formation of low mass stars and brown dwarfs.

\end{abstract}

\keywords{instrumentation: adaptive optics --- binaries: general --- stars: evolution --- stars: formation
--- stars: low-mass,    brown dwarfs}

\section {Introduction}

The detailed formation of stars is still a poorly understood process. In
particular, the formation of the lowest mass stars and brown dwarfs is
uncertain. Detailed 3D simulations of star formation by \cite{bat02}
suggest that stellar embryos form into ``mini-clusters'' which
dynamically decay ``ejecting'' the lowest mass members. Such theories
can explain why there are far more field brown dwarfs (BD) compared to
BD companions of solar type stars \citep{mcc03} or early M stars
\citep{hin02}. Moreover, these theories which invoke some sort of
dynamical decay \citep{dur01} or ejection \citep{rep02} suggest that
there should be no wide ($>20$ AU) very low mass (VLM; $M_{tot}<0.185
M_{\sun}$) binary systems observed. Indeed, the AO surveys of
\cite{clo03a} and the HST surveys of \cite{rei01a,bur03,bou03,giz03} have not
discovered any wide ($>16$ AU) VLM systems of the 34 systems known
to date. As well, the dynamical biasing towards the ejection of the
lowest mass members naturally suggests that the frequency of VLM
binaries should be much less ($\la 5\%$ for $M_{tot}\sim 0.16
M_{\sun}$) than for more massive binaries ($\sim 60\%$ for $M_{tot}
\sim 1 M_{\sun}$). Indeed, observations suggest that the binarity of
VLM systems with $M_{tot} \la 0.185 M_{\sun}$ is $10-15\%$
\citep{clo03a, bur03} which, although higher than predicted is still
lower than that of the $\sim 60\%$ of G star binaries \cite{duq91}.

Despite the success of these decay or ejection scenarios in predicting
the observed properties of binary stars, it is still not clear that
``mini-clusters'' even exist in the early stages of star
formation. To better understand whether such ``mini-clusters'' do
exist we have examined the closest major OB star formation cluster for
signs of such mini-clusters. Here we focus on the $\theta^1$ Ori stars
in the Trapezium cluster. Trying to determine if some of the tight
star groups in the Trapezium cluster are gravitationally bound is a
first step to determining if bound ``mini-clusters'' exist. In
particular, we will examine the case of the $\theta ^1$ Ori B and A
groups.

The Trapezium OB stars ($\theta^1$ Ori A, B, C, D, and E) consists of
the most massive OB stars located at the center of the Orion Nebula
star formation cluster (for a review see \cite{gen89}). Due to the
luminous nature of these stars they have been the target of several
high-resolution imaging studies. Utilizing only tip-tilt compensation
\cite{mcc94} mapped the region at $K^\prime$ from the 3.5-m Calar Alto
telescope. They noted that $\theta^1$ Ori B was really composed of 2
components ($B_1$ \& $B_2$) about $\sim 1\arcsec$ apart. Higher $\sim
0.15\arcsec$ resolutions were obtained from the same telescope by
\cite{pet98} with speckle holographic observations. At these higher
resolutions \cite{pet98} discovered that $\theta^1$ Ori $B_2$ was
really itself a $0.1\arcsec$ system ($B_2$ \& $B_3$) and that
$\theta^1$ Ori A was really a $\sim0.2\arcsec$ binary ($A_1$ \&
$A_2$). A large AO survey of the inner 6 square arcminutes was carried
out by \cite{sim99}, who discovered a very faint (100 times fainter
than $B_1$) object ($B_4$) located just $0.6\arcsec$ between $B_1$ and
$B_2$. Moreover, a spectroscopic survey \citep{abt91} showed that
$B_1$ was really an eclipsing spectroscopic binary ($B_1$ \& $B_5$;
sep. 0.13 AU; period 6.47 days). As well, $\theta^1$ Ori $A_1$ was
also found to be a spectroscopic binary ($A_1$ \& $A_3$; sep. 1 AU;
\cite{bos89} ). \cite{wei99} carried out bispectrum speckle
interferometric observations at the larger Russian SAO 6-m telescope
(2 runs in 1997 and 1998). These observations showed $\theta^1$ Ori C
was a very tight 0.033$\arcsec$ binary. These observations also
provided the first set of accurate relative positions for these
stars. \cite{sch03} has continued to monitor this cluster of stars and
has independently detected an orbital motion (of $\Delta PA \sim
6^{\circ}$ for $\theta^{1}$ Ori $A_2$ around $A_1$ and a $\Delta PA$
of $\sim 8^{\circ}$ for $\theta^{1}$ Ori $B_3$ around $B_2$ over a 5.5 yr
baseline). They conclude that this is real orbital motion. We present
additional recent AO observations of these binaries as an independent
check to confirm that these motions are indeed real.

We first utilized the Gemini telescope (with the
Hokupa'a AO system) and then observed $\theta^1$ Ori B during
commissioning of the world's first secondary deformable mirror at the
6.5-m MMT telescope. In this paper we outline how the observations
were carried out, and how the stellar positions were measured. We fit
the observed positions to calculate velocities (or upper limits) for
the $\theta^1$ Ori B \& A stars. In agreement with \cite{sch03}, we find that there is good evidence
that the $\theta^1$ Ori B group may be a bound ``mini-cluster'' and
that the $\theta^1$ Ori A group is also likely gravitationally bound.

\section{OBSERVATIONS}

 We have utilized the University of Arizona adaptive secondary AO
 system to obtain the most recent high resolution images of the young
 stars in the Trapezium cluster (the $\theta^{1}$ Ori group).

\subsection{The World's First Adaptive Secondary AO System Scientific Results}

The 6.5 m MMT telescope has a unique adaptive optics system. To reduce
the aberrations caused by atmospheric turbulence all AO systems have a
deformable mirror which is updated in shape at $\sim 500$ Hz. Until
now all adaptive optics systems have located this deformable mirror
(DM) at a re-imaged pupil (effectively a compressed image of the
primary mirror). To reimage the pupil onto a DM typically requires 6-8
warm additional optical surfaces which significantly increases the
thermal background and decreases the optical throughput of the system
\citep{llo00}. However, the MMT utilizes a completely new type of DM.
 This DM is both the secondary mirror of the
telescope and the DM of the AO system. 
 In this
manner there are no additional optics required in front of the science
camera. Hence the emissivity is lower and the possibility of thermal IR AO imaging \citep{clo03b,bil03} becomes a reality.

The DM consists of 336 voice coil actuators that push on 336 small
magnets glued to the backsurface of a thin (2.0 mm thick) 642 mm
aspheric ULE glass ``shell'' (for a detailed review of the secondary mirror see \citep{bru03a,bru03b}).
 We have complete
positional control of the surface of this reflective shell by use of
a capacitive sensor feedback loop. This positional feedback loop
allows one to position an actuator of the shell to within 4 nm rms (total surface
errors amount to only 40 nm rms over the whole secondary). The AO
system samples at 550 Hz using 108 active subapertures. For a detailed
review of the MMT AO system see
\cite{wil03,wil03b} and references within. 

\subsection{MMT AO Observations}

\begin{figure}
\includegraphics[angle=0,width=\columnwidth]{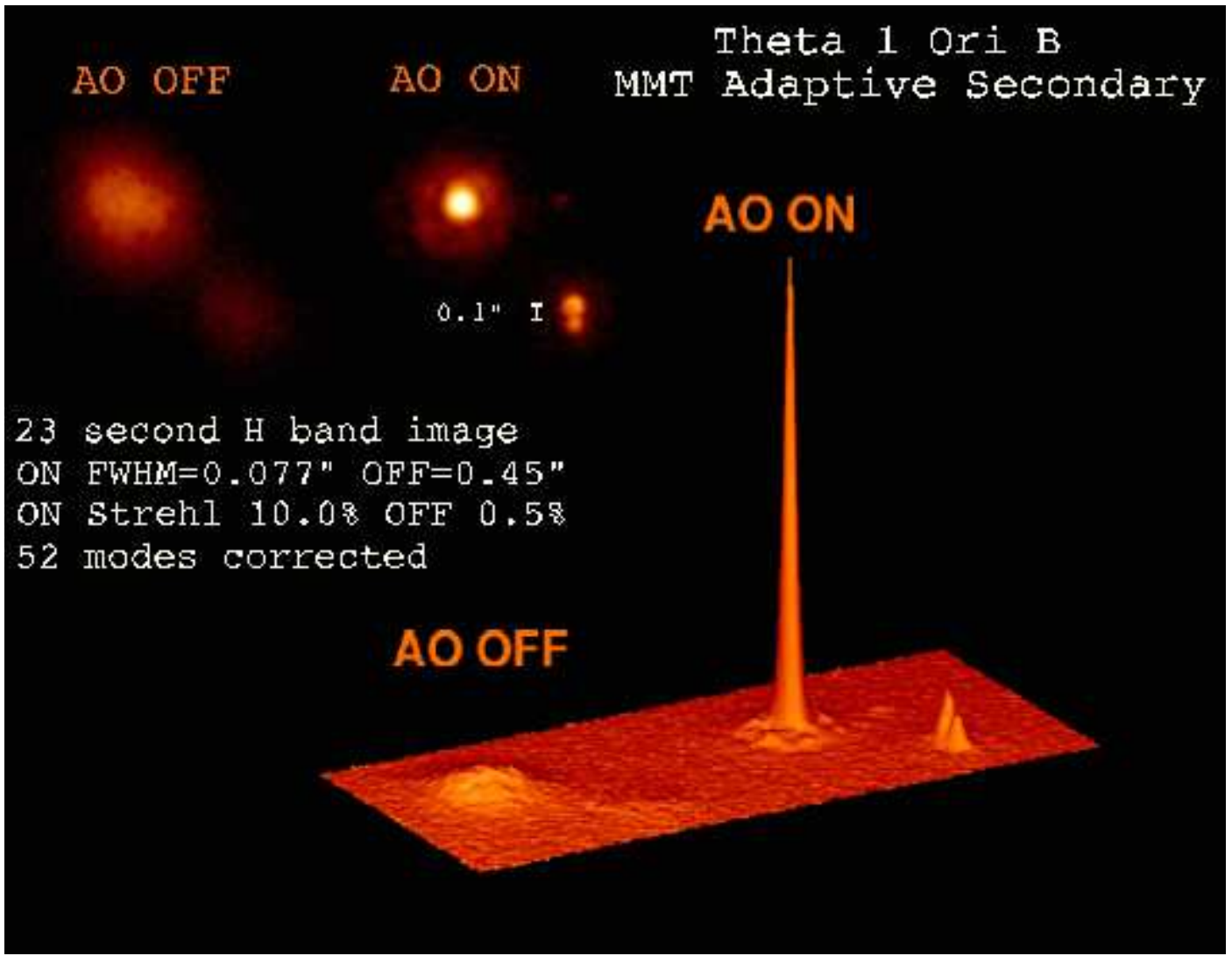}\caption{
A typical example of how the the Adaptive Optics (AO) system can make
very sharp images. With AO "OFF" $\theta^1$ Ori B appears to be just 2
stars. With AO turned "ON" it is clearly a tight group of 4 visual
stars. Note how with AO correction the peak intensity increases by 20
times and the resolution becomes ten times better.}
\label{fig2}
\end{figure}

During our second engineering run we observed the $\theta^{1}$ Ori B
group on the night of Jan 20, 2003 (UT). The AO system corrected the
lowest 52 system modes and was updated at 550 Hz. The closed loop
bandwidth was estimated at 30 Hz 0 dB. Without AO correction our
images had FWHM=$0.6\arcsec$, after AO correction our 23 second images
had improved to FWHM=$0.070\arcsec$ (close to the diffraction limit of
$0.056\arcsec$ in the H-band). A detailed analysis suggested that
during our engineering run a 40 Hz vibration in the MMT telescope
increased our FWHM by $\sim 0.015\arcsec$ and decreased our Strehl by
a factor of two. We are in the process of identifying and decreasing
the effect of this 40 Hz vibration. In any case, as Figure \ref{fig2}
clearly shows, there is a large improvement in image quality (the
Strehl increases by 20 times) with the adaptive secondary AO system.

\subsubsection{ The Indigo Near-IR Video Camera}

Since these observations were carried out during the engineering run
we utilized a commercially available 320x256 InGaAs 0.9-1.68 $\mu m$
``Merlin-NIR'' video camera. Although this commercial camera (produced by the
Indigo company) is not nearly as sensitive as our facility AO camera
(AIRES; \cite{mcc98}) it still provides excellent dynamical
information about the performance of the AO system on bright objects
(it will be replaced by the ARIES camera in the fall of 2003). Here we use it as a simple NIR (H band) science camera.

\begin{figure}
\includegraphics[angle=0,width=\columnwidth]{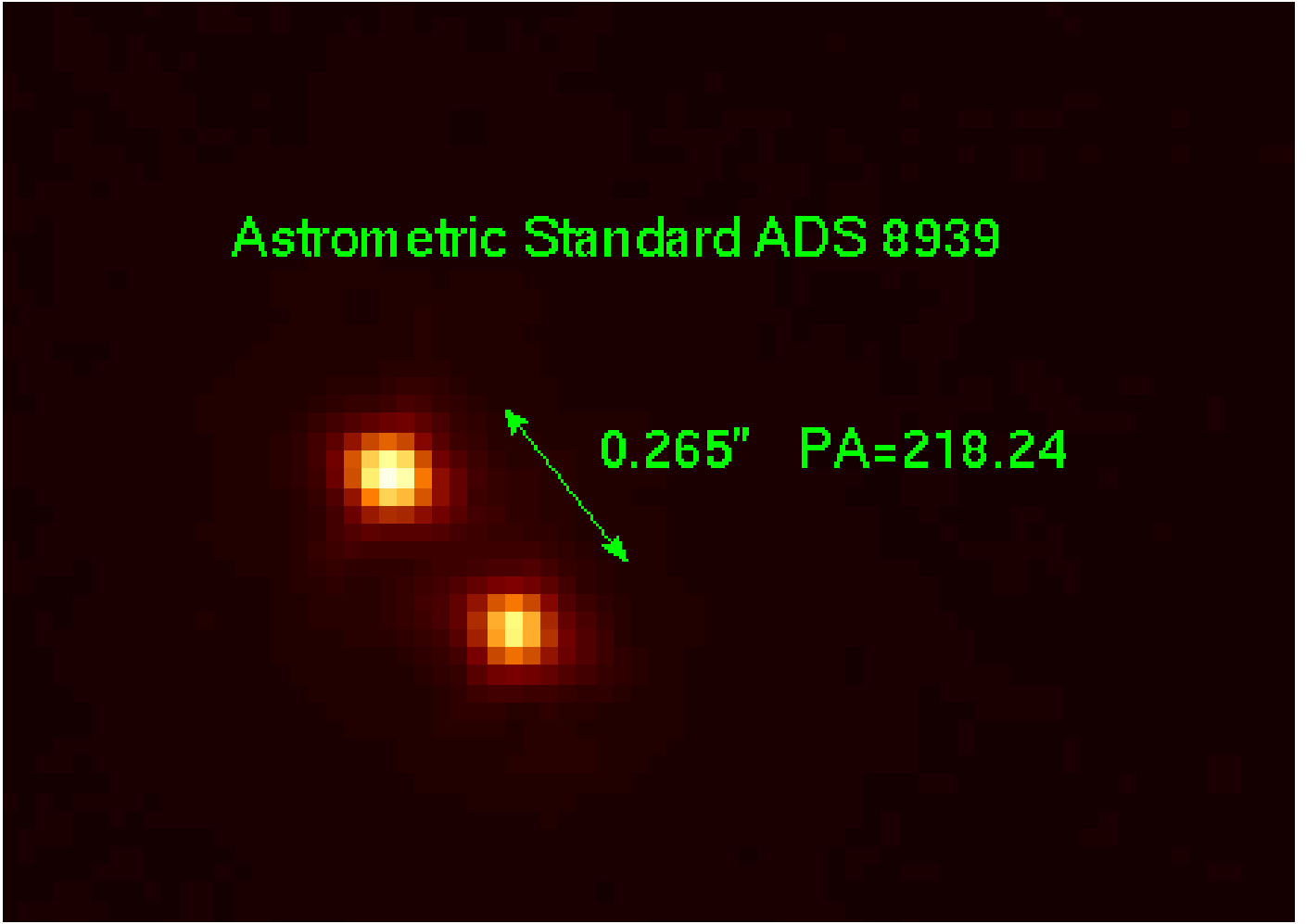}
\caption{
An H band MMT AO image of the astrometric binary ADS 8939 (WDS
13329+3454; STT 269AB). The well known orbit (WDS Grade level 2) of
this binary star predicted a separation of $0.265\arcsec$ and a PA of
$218.237^{\circ}$ for UT Jan 19, 2003 (the night of this
observation). For these values we derived that the Indigo camera had a
platescale of $0.0242\arcsec$/pixel. This 10 second integration had a
mid-point time of UT 12:21:30, hence the parallactic angle during this
exposure was $-107.6^{\circ}$. Rotating the image by $-107.6^{\circ}$
(clockwise) resulted in a measured PA of $218.35^{\circ}$ which
indicates North is $0.113^{\circ}$ east of the Indigo's Y axis. Linear
color scale. North is up and east is left.}
\label{fig2a}
\end{figure}

\begin{figure}
\includegraphics[angle=0,width=\columnwidth]{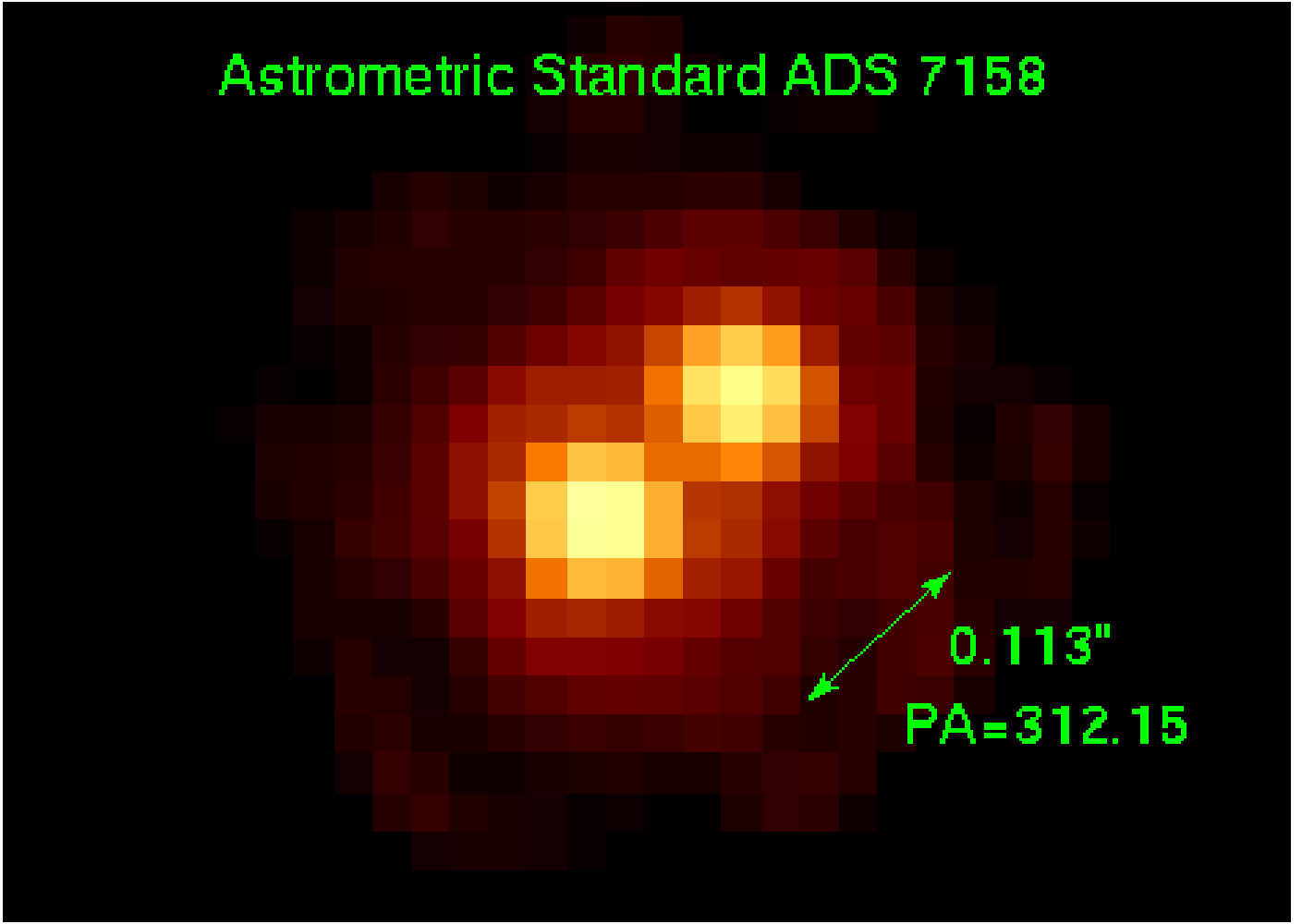}
\caption{
An H band image of the astrometric binary ADS 7158 (WDS 09036+4709; A
1585). The well known orbit (WDS Grade level 2) of this binary
star predicted a separation of $0.111\arcsec$ and a PA of
$312.764^{\circ}$ for UT Jan 20, 2003 (the night of this
observation). We utilized these values to check the $0.0242\arcsec$/pixel
platescale and orientation (north being $0.113^{\circ}$ east of the
Indigo's Y axis) that were obtained from the ADS 8939 observations for
the Indigo camera (see Figure \ref{fig2a}). The above 10 second integration
had a mid point time of UT 8:18:50, hence the parallactic angle during
this exposure was $-171.0^{\circ}$. Rotating the image by
$-171.0^{\circ}$ (and correcting for the $0.113^{\circ}$ misalignment
of the Y axis) resulted in a measured PA of $312.146^{\circ}$ which
which is incorrect by $0.62^{\circ}$. Hence we conservatively estimate
our PA is calibrated to with $\pm1^{\circ}$. The separation of ADS 7158
is 4.677 pixels suggesting a platescale of
$0.0241\arcsec$/pixel. Hence we estimate a conservative
$\pm0.002\arcsec$ error in the Indigo platescale of
$0.0242\arcsec$/pixel. Logarithmic color scale, note the Airy rings
around each component. North is up and east is left.}
\label{fig2b}
\end{figure}

The Indigo camera was fed by a relay lens that converted the f/15 AO
corrected beam to a f/39 beam yielding $0.0242\pm0.0020\arcsec$ per
$30\mu m$ pixel (providing a $7.7\times 6.2\arcsec$ FOV). Astrometric
standards ADS 8939 and ADS 7158 were observed to calibrate this
platescale and error (see Figures \ref{fig2a} \& \ref{fig2b}). It was
found that the direction of north was slightly ($0.113^{\circ}$) east
of Indigo's Y axis (when the parallactic angle was zero (transit) and
one is looking towards the south). During this commissioning run we
did not observe with the MMT Cassegrain derotator tracking field
rotation, hence all images must be rotated by the appropriate
parallactic angle (plus $0.113^{\circ}$) to have north up and east to
the left on the Indigo camera.

The camera was mounted under a high optical quality dichroic
which sent the visible light (0.5-1 $\mu m$) to the 108 subaperture
shack-Hartmann wavefront sensor (WFS). The infrared light ($\lambda >
1 \mu m$) was transmitted to the Indigo camera. The camera had a
standard H band filter ($1.6\mu m$) mounted 3 inches from focus in a
light-tight barrel.

To maximize the sensitivity of the Indigo camera we carried out a
standard ``2-point'' calibration on a both a dark (cold) flat field
source and on a bright (hot) source to scale the automatic gain
control/dynamic range of the camera's electronics. This appeared to
yield images that were auto flat fielded to a few percent in accuracy
when the counts were between the linear range defined by the dark and
bright calibration flats. The camera was remotely controlled via a
serial port. Digital (16 bit) data were streamed to the control PC's
hard drive. Data could be acquired as fast as 50 frames per second
(although data in this paper was acquired at 15 frames/sec to sample
longer periods on the sky). Integration times can range from 1-16000
$\mu s$. The lack of a longer integration time (since the camera is
primarily intended for commercial high-background, high-bandwidth
applications) leads to most sources being read-noise limited. However,
we found that point sources of $H\sim11$ could be detected in 3 s of
total exposure (200 16 ms frames) with AO correction at the
MMT. Although insensitive by most astronomical standards the Indigo
camera is able to capture temporal events of durations as short as 1
$\mu s$. In this paper we will focus on the ability of the Indigo
camera to produce high resolution ($0.07\arcsec$) images of the
$\theta^{1}$ Ori B group.

\subsubsection{Reducing the Indigo MMT AO Data}

\begin{figure}
\includegraphics[angle=0,width=\columnwidth]{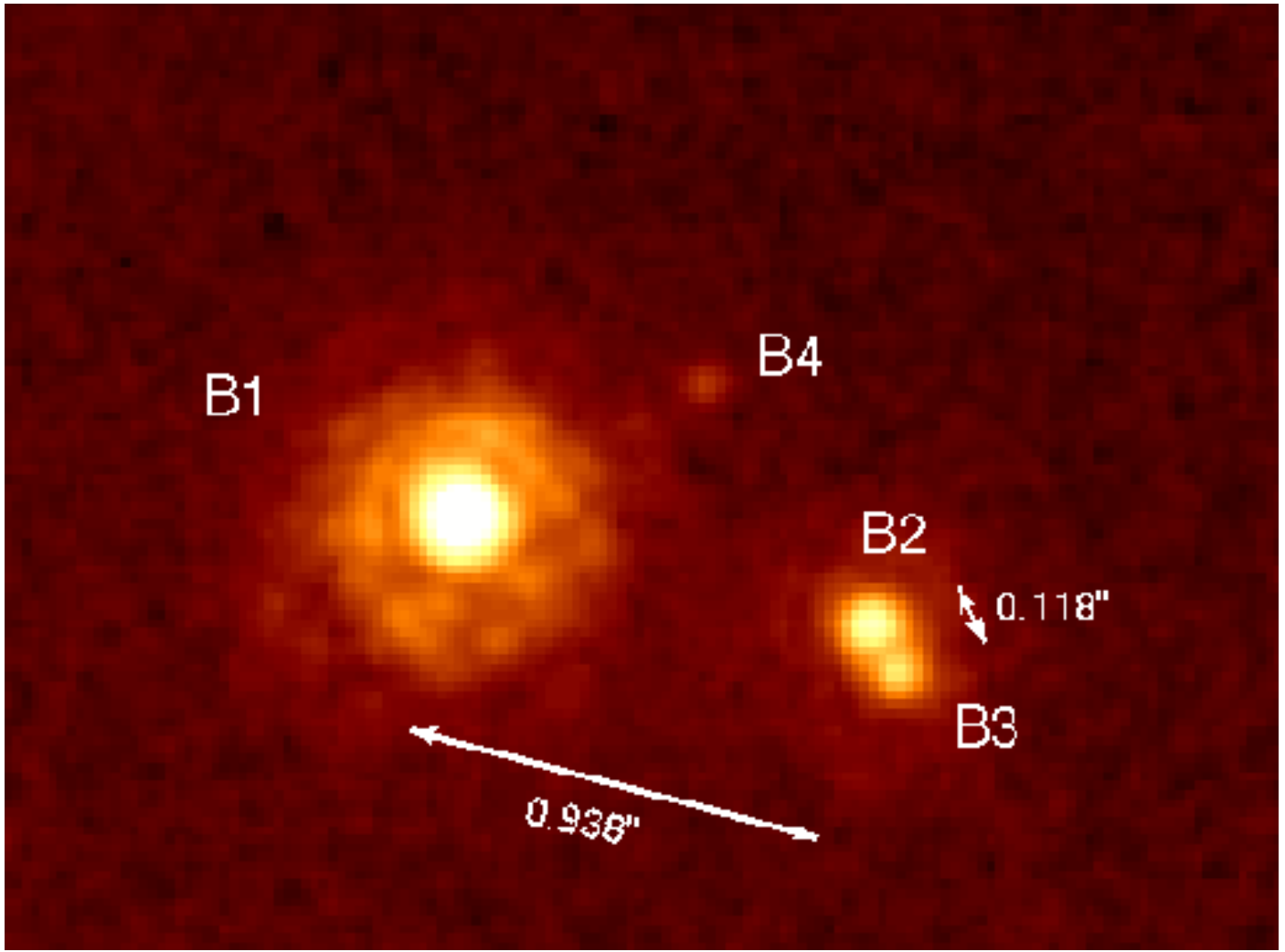}
\caption{
Detail of the $\theta^{1}$ Ori B group as imaged at $0.077\arcsec$
resolution (in the H band) with the MMT AO system and the Indigo IR
camera. Logarithmic color scale. North is up and east is left. Note that the object ``$B_1$'' is really an eclipsing spectroscopic binary ($B_1B_5$); where the unseen companion $B_5$ orbits $B_1$ every 6.47 days \citep{abt91}.}
\label{fig3}
\end{figure}

For the $\theta^{1}$ Ori B group we obtained 7 series of 200x16 ms
data cubes with the Indigo camera. The data from each cube was simply
averaged together to produce 7 individual 3.2 second exposures. A
similarly reduced cube of ``sky'' images was subtracted from each data
set. These 7 sky-subtracted exposures were then rotated (in IRAF) by
the current parallactic angle (plus the $0.113^{\circ}$ offset) so
north was aligned with the Y axis, and east is the negative X
axis. Then each of the 7 images were cross-correlated and aligned with a
cubic spline interpolator. Then the final stack of images were median
combined to produce the final image. The final image is displayed in
Figure
\ref{fig3}.

\subsection{Hokupa'a/Gemini Images of the Trapezium}

In addition to our excellent MMT images of the $\theta^{1}$ Ori B
group we also have an epoch of $K^{\prime}$ images of the central
$30\arcsec$ of the cluster. These Hokupa'a/Gemini \citep{gra98,clo98a}
AO images were taken September 19, 2001. We acquired a series of 10 short
(1 s) images and dithered the telescope in a 10x10$\arcsec$ box while
AO guiding off $\theta^{1}$ Ori B itself (as in the case of the MMT AO
observations). We utilized the QUIRC IR camera \citep{hod96} with a
calibrated platescale of $0.0199\pm0.0002\arcsec$/pix \citep{pot02a}.

\subsubsection{Reducing the Gemini data}

We have developed an AO data reduction pipeline in the IRAF language
which maximizes sensitivity and image resolution. This pipeline is
standard IR AO data reduction and is described in detail in
\cite{clo02a,clo02b}.  

\begin{figure}
\includegraphics[angle=0,width=\columnwidth]{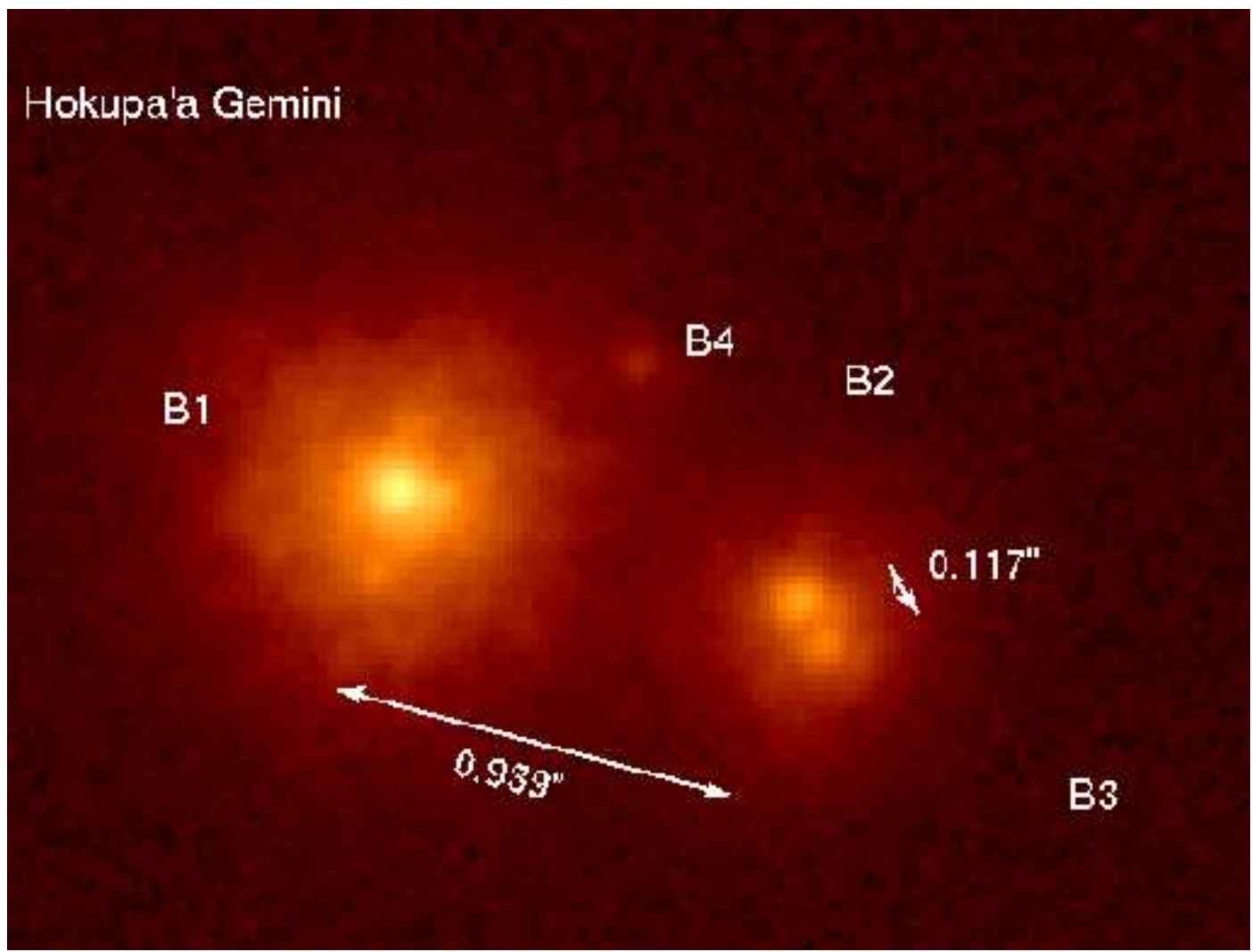}\caption{
The Gemini/Hokupa'a images of the $\theta^{1}$ Ori B group in the $K^\prime$ band. Resolution $0.085\arcsec$. Logarithmic color scale. North is up and east is left. 
}
\label{fig4}
\end{figure}

 The pipeline cross-correlates and aligns each image, then rotates
each image so north is up (to an accuracy of $\pm0.3$ degrees) and
east is to the left, then median combines the data with an average
sigma clip rejection at the $\pm2.5 \sigma$ level. By use of a
cubic-spline interpolator the script preserves image resolution to the
$<0.02$ pixel level. Next the custom IRAF script produces two final
output images, one that combines all the images taken (see Figure \ref{fig4}) and another
where only the sharpest 50\% of the images are combined (this high-Strehl image was very similar to that shown in Figure \ref{fig4}, just a bit noisier -- and so was not further analyzed).

\begin{figure}
\includegraphics[angle=0,width=\columnwidth]{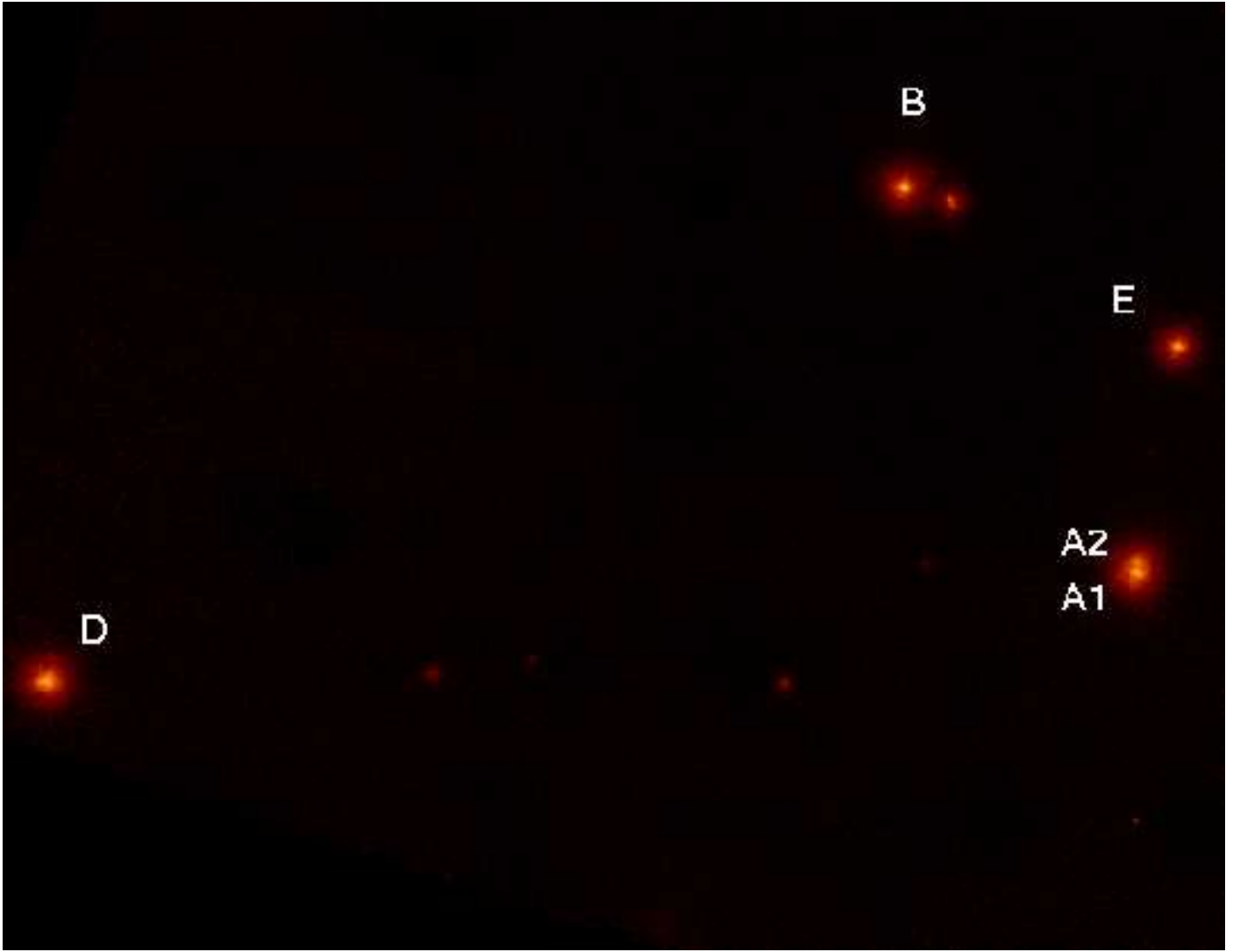}\caption{
The upper part of the  $\theta^{1}$ Ori cluster as imaged over $30\times 30\arcsec$ FOV at Gemini with the Hokupa'a AO system. Logarithmic color scale. North is up and east is left. Note that the object ``$A_1$'' is really a spectroscopic binary ($A_1A_3$); where the unseen companion $A_3$ is separated from $A_1$ by 1 AU \citep{bos89}
}
\label{fig5}
\end{figure}

The final image (see Figures \ref{fig4} and \ref{fig5}) has
FWHM=$0.085\arcsec$ which is just slightly worse than the MMT
data. Even though Gemini is a larger telescope (8.2-m), Hokupa'a's
fitting error (36 elements over 50 meters$^2$) is worse than that of
the MMT (52 modes over 33 meters$^2$), hence higher resolution images
can result from the smaller of the two telescopes (Gemini has a
diffraction-limit of $0.056\arcsec$ at $K^{\prime}$ similar to that of
the MMT at H). However, Hokupa'a's curvature WFS could guide on much
fainter (R$\sim17$) guide stars
\citep{clo02a,clo02b,sie02a}.

\section{Reductions}

In Table \ref{tbl-1} we present the analysis of our MMT and Gemini
images in Figures \ref{fig3} and \ref{fig4}. The photometry was based
on DAOPHOT's PSF fitting photometry task ALLSTARS \citep{ste87}. The
PSF used was $\theta^{1}$ Ori $B_{1}$ itself. Since all the members of
the $\theta^{1}$ Ori $B$ group are located within $1\arcsec$ of
$\theta^{1}$ Ori $B_{1}$ the PSF fit is excellent (there is no
detectable change in PSF morphology due to anisoplanatic effects
inside the $\theta^1$ Ori B group \citep{dio00}).

\begin{figure}
\includegraphics[angle=90,width=\columnwidth]{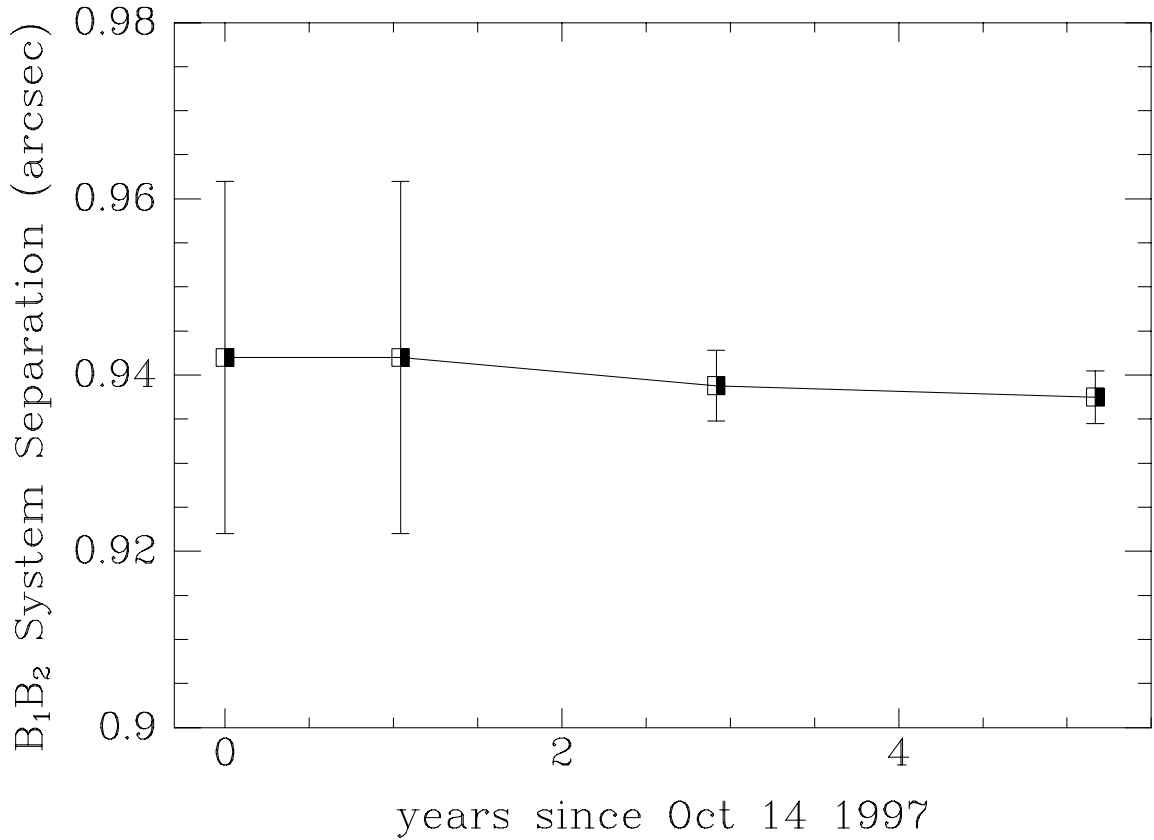}\caption{
The separation between $\theta^{1}$ Ori $B_1$ and $B_2$. Note how over
five years of observation there has been little significant relative
proper motion observed (-0.0006$\pm0.0019\arcsec$/yr; which is
insignificantly different from a constant). If the group is
gravitationally bound the separation should be roughly constant over
five years. The observed rms scatter from a constant value is indeed a mere
$\pm0.0019\arcsec$, suggesting the whole $\theta^{1}$ Ori B group is
likely physically bound together. The first 2 data points are speckle
observations from the 6-m SAO telescope \citep{wei99}, the next point
is from our Gemini/Hokupa'a observations and the last data point is
from the MMT AO observations. }
\label{fig6}
\end{figure}

\begin{figure}
\includegraphics[angle=90,width=\columnwidth]{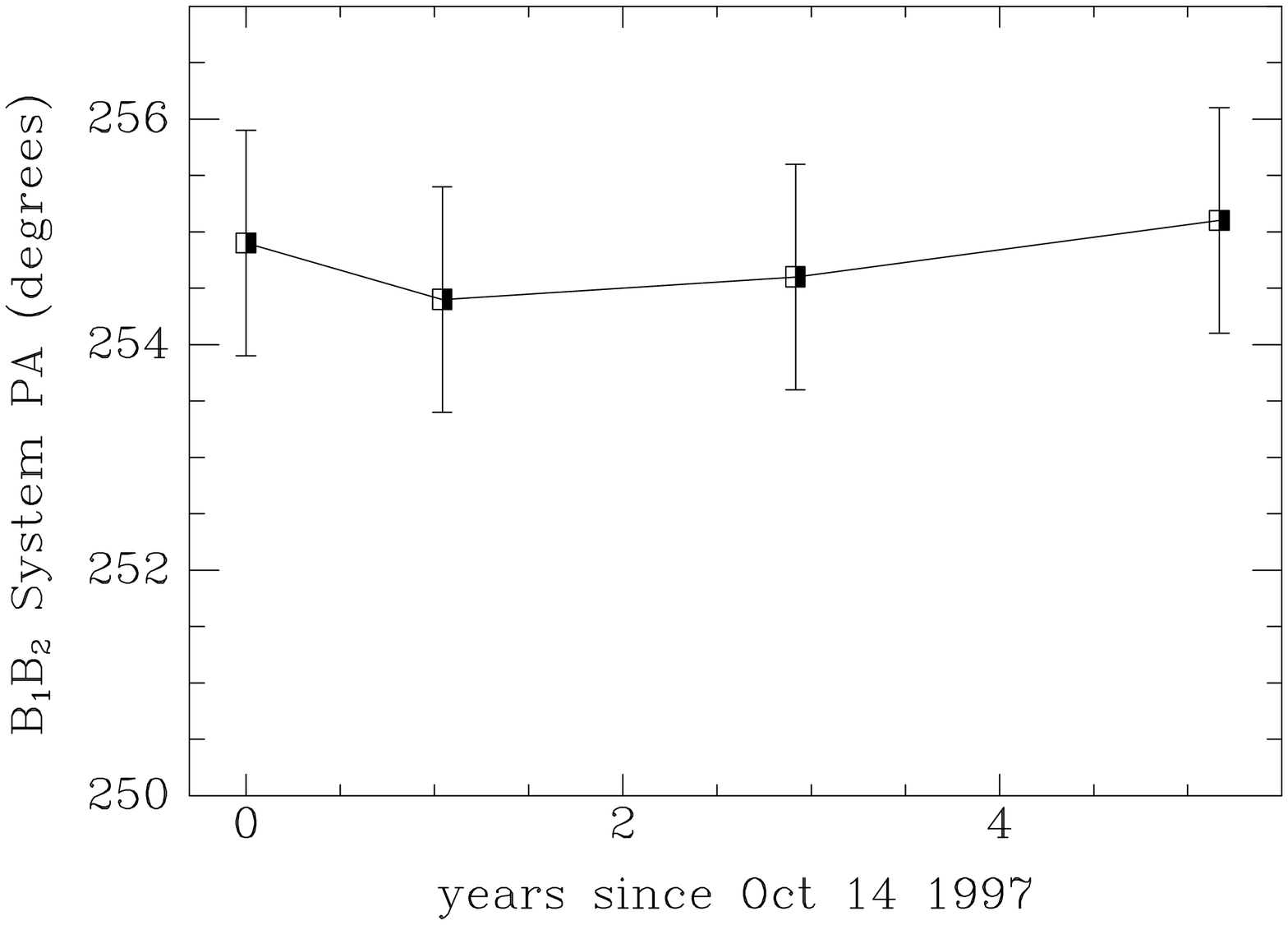}\caption{
The position angle between $\theta^{1}$ Ori $B_1$ and $B_2$. Note how
over five years of observation there has been no significant relative
proper motion observed (0.07$\pm0.25^\circ$/yr which is insignificantly different from a constant). The
error from a constant value is a mere $\pm0.3^\circ$.The first 2 data
points are speckle observations from the 6-m SAO telescope
\citep{wei99}, the next point is from our Gemini/Hokupa'a observations
and the last data point is from the MMT AO observations. }
\label{fig7}
\end{figure}

\begin{figure}
\includegraphics[angle=90,width=\columnwidth]{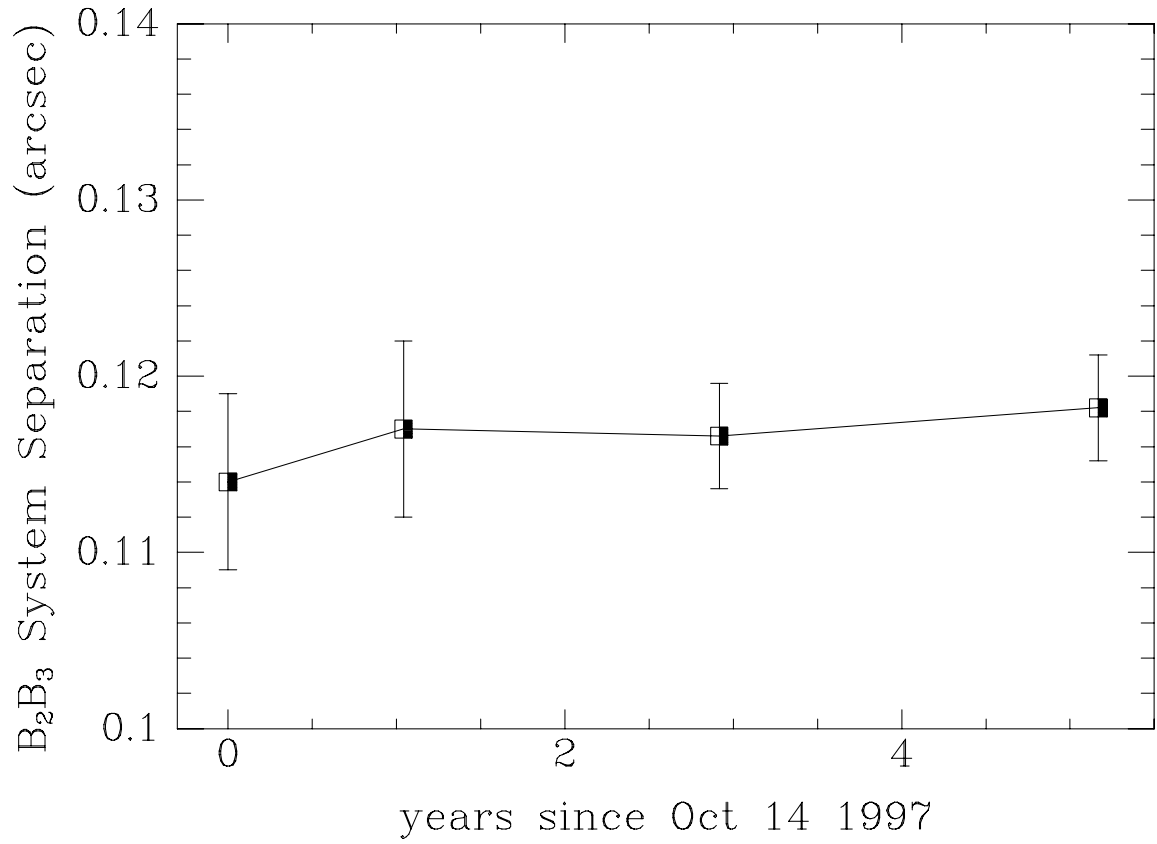}\caption{
The separation between $\theta^{1}$ Ori $B_2$ and $B_3$. Note the lack
of any significant relative motion ($0.0006\pm0.0010\arcsec$/yr). The rms scatter from a constant value is only $0.001\arcsec$. There appears to very little change in the separation of the $B_2B_3$ system. The
first 2 data points are speckle observations from the 6-m SAO
telescope \citep{wei99}, the next point is from our Gemini/Hokupa'a
observations and the last data point is from the MMT AO
observations. }
\label{fig8}
\end{figure}

Since the PSF model was so accurate and the data had such high signal
to noise (and high resolution) it was possible for DAOPHOT to measure
relative positions to within $0.003\arcsec$. We estimate this error based on
the scatter of the $\theta^{1}$ Ori $B_{1}B_{2}$ separation (which
should be very close to a constant since the $B_{1}B_{2}$ system has
an orbital period of $\sim 2000$ yr). The lack of any motion between $B_1$ and $B_2$ is also confirmed by \cite{sch03}. Our data is summarized in Table
\ref{tbl-1}. Linear (weighted) fits to the
data in Table \ref{tbl-1} (Figures \ref{fig6} to \ref{fig11})
yield the velocities shown in Table \ref{tbl-1}. The overall error in
the relative proper motions observed is $\sim 0.002\arcsec$/yr in
proper motion ($\sim 4$ km/s).

\section{ANALYSIS}

\begin{figure}
\includegraphics[angle=90,width=\columnwidth]{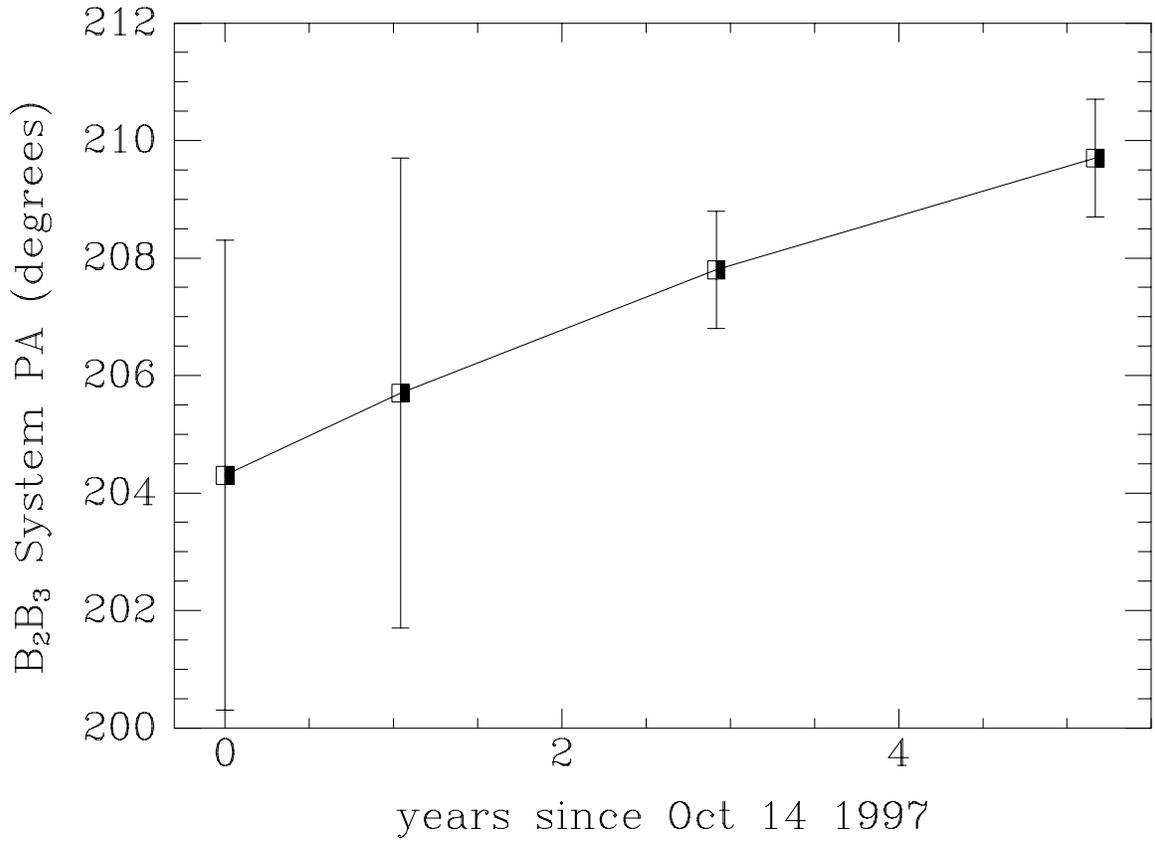}\caption{
The position angle of $\theta^{1}$ Ori $B_2$ and $B_3$. Here we
observe what may be real orbital motion of $B_3$ moving
counter-clockwise (at 0.93$\pm0.49^\circ$/yr; correlation significant at the
99.2\% level) around $B_2$. This small amount of motion is consistant
with the $B_2B_3$ system being bound. The first 2 data points are
speckle observations from the 6-m SAO telescope \citep{wei99}, the
next point is from our Gemini/Hokupa'a observations and the last data
point is from the MMT AO observations. }
\label{fig9}
\end{figure}

\begin{figure}
\includegraphics[angle=90,width=\columnwidth]{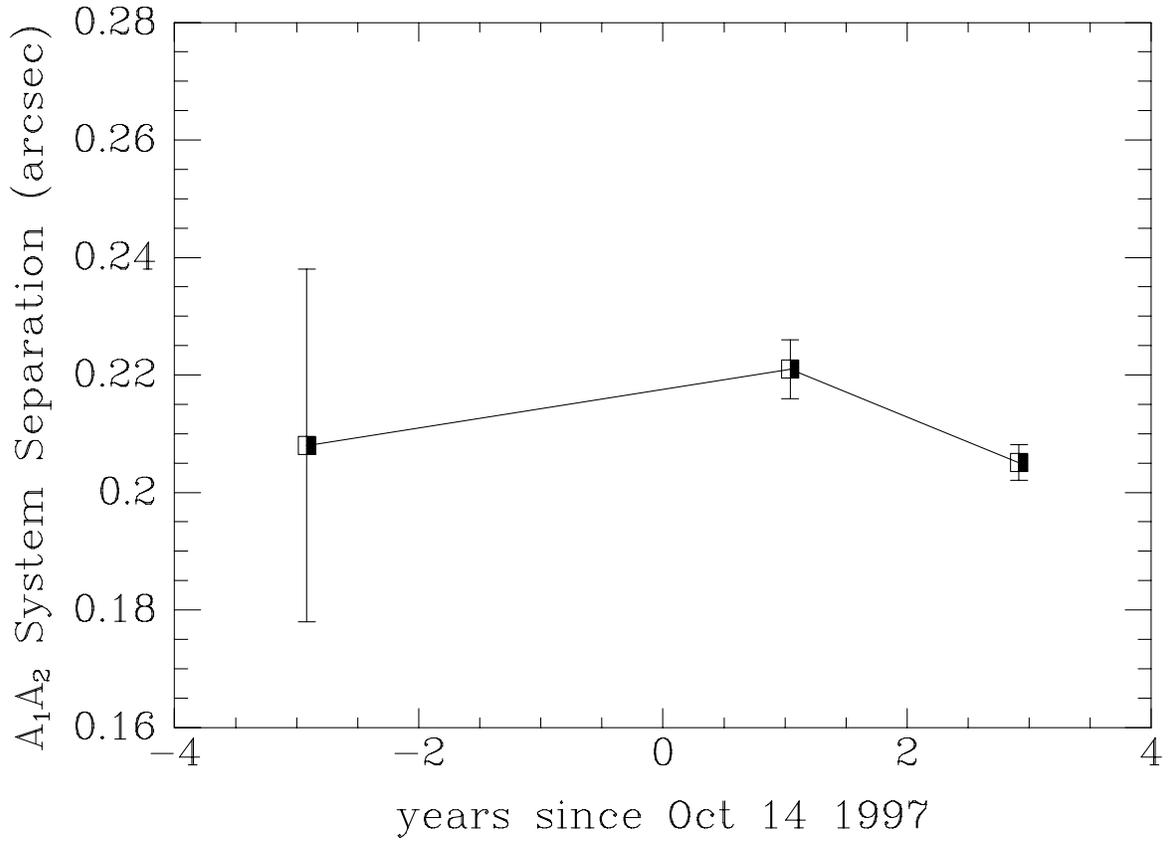}\caption{
The separation between $\theta^{1}$ Ori $A_1$ and $A_2$. There is a small negative changes in the orbital separation
($-0.0064\pm0.0027\arcsec$/yr) as $A_2$ moves towards $A_1$. The first data point is from speckle observations
at the 3.5-m Calar Alto telescope \citep{pet98}, the next point is from
a speckle observation from the 6-m SAO telescope
\citep{wei99}, the last point is from our Gemini/Hokupa'a
observations. }
\label{fig10}
\end{figure}

\begin{figure}
\includegraphics[angle=90,width=\columnwidth]{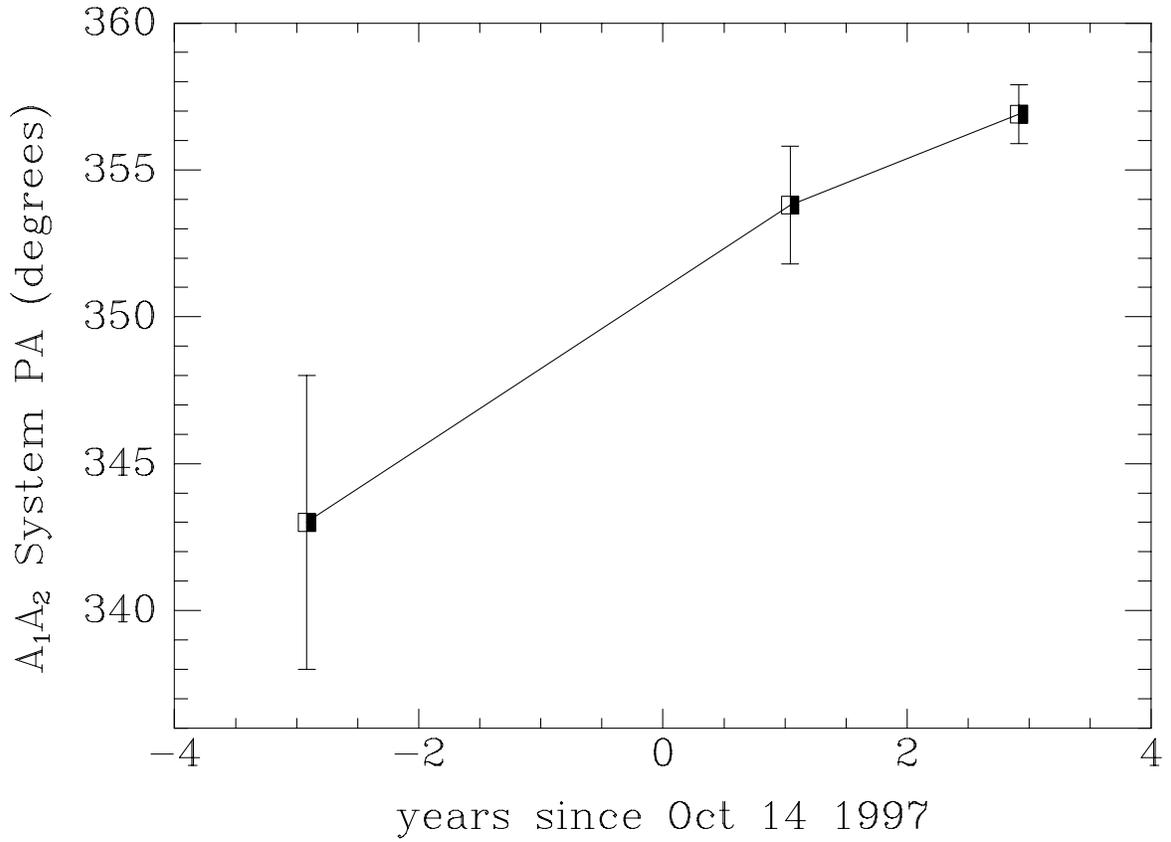}\caption{
The position angle of $\theta^{1}$ Ori $A_1$ and $A_2$. There does
appear to be significant changes in the position angle as $A_2$ moves
counter clockwise (at 2.13$\pm0.73^\circ$/yr) around $A_1$. This
relatively small motion is consistent with the $A_1A_2$ system being
bound. The first data point is from speckle observations at the 3.5-m
Calar Alto telescope \citep{pet98}, the next point is from a speckle
observation from the 6-m SAO telescope \citep{wei99}, the last point
is from our Gemini/Hokupa'a observations. }
\label{fig11}
\end{figure}

With these accuracies it is now possible to determine whether these
stars in the $\theta^{1}$ Ori $B$ group are bound together, or merely
chance projections in this very crowded region. As can be seen from
Table \ref{tbl-1} and Figures \ref{fig6} -- \ref{fig11} there is very
little relative motion between any of the members of the $\theta^{1}$
Ori $B$ group. Therefore it is possible that the group is physically bound
together.


If we adopt the masses of each star from the \cite{sie97,ber96} tracks
fit by \cite{wei99} we find masses of: $B_1\sim7M_{\sun} $;
$B_2\sim3M_{\sun} $; $B_3\sim2.5M_{\sun} $; $B_4\sim0.2M_{\sun} $;
$B_5\sim 7 M_{\sun}$; $A_1\sim20M_{\sun} $; $A_2\sim 4M_{\sun} $; and
$A_3\sim 2.6 M_{\sun}$. Based on these masses (which are similar to those adopted by \cite{sch03}) we can comment on
whether the observed motions are less than the escape velocities
expected for simple face-on circular orbits.

Our combination of high spatial resolution and high signal to noise
yields an error in the proper motions of only $\sim 0.002\arcsec$/yr
according to the scatter in the $B_1B_2$ and $B_1B_3$ systems (see
Table \ref{tbl-1}). We have observed orbital motion in the very tight
$\theta^{1}$ Ori $B_{2}B_{3}$ (see Figure \ref{fig9}) and $\theta^{1}$
Ori $A_{1}A_{2}$ (see Figure \ref{fig11}) systems, with 52 and 94 AU
separations; respectively.

\subsection{Is the $\theta^1$ Ori $B_2B_3$ System Physical?}

The relative velocity in the $\theta^{1}$ Ori $B_{2}B_{3}$ system (in
the plane of the sky) is $\sim 4.2\pm2.1$ km/s (mainly in the azimuthal direction; see Figure \ref{fig9}). This is a reasonable
$V_{tan}$ since an orbital velocity of $\sim 6.7$ km/s is expected
from a face-on circular orbit from a $\sim5.5 M_{\sun}$ binary system
like $\theta^{1}$ Ori $B_{2}B_{3}$ with a 52 AU projected
separation. It is worth noting that this velocity is also greater than
the $\sim 3$ km/s \cite{hil98} dispersion velocity of the cluster.
Hence it is most likely that these two $K^\prime =7.6$ and $K^\prime
=8.6$ stars (separated by just 0.116$\arcsec$) are indeed in orbit around each other. Moreover, there are only 10 stars known to have $K^\prime < 8.6$ in the inner
$30\times 30\arcsec$ (see Figure \ref{fig5}), we
can estimate that the chances of finding two bright ($K^\prime < 8.6$)
stars within $0.116\arcsec$ is a small $<10^{-4}$ probability.

Our observed velocity of $0.93\pm0.49^{\circ}$/yr is consistent (in both direction and magnitude) with
the $1.4^{\circ}$/yr observed by \cite{sch03}. This suggests that the
AO and speckle datasets are both detecting real motion. Moreover, since
this motion is primarily azimuthal strongly suggests an orbital arc of $B_3$
orbiting $B_2$.

\subsection{Is the $\theta^1$ Ori $A_1A_2$ System Physical?}

We observe $\sim 16.5\pm5.7$ km/s of relative motion in the
$\theta^{1}$ Ori $A_{1}A_{2}$ system (mainly in the azimuthal direction; see Figure \ref{fig11}). This is higher than the average
dispersion velocity of $\sim 3$ km/s but still close to an estimated
periastron velocity of the $\sim 20 M_{\sun}$ $A_{1}A_{2}$ system
(projected separation of 94 AU). Hence it is highly likely that these
two $K^\prime =6.0$ and $K^\prime =7.6$ stars (separated by just
0.21$\arcsec$) are indeed in orbit around each other. In addition, there are only 8 stars known to have $K^\prime < 7.6$ in the inner
$30\times 30\arcsec$ (see Figure \ref{fig5}), we
can estimate that the chances of finding two bright ($K^\prime < 8.6$)
stars within $0.21\arcsec$ is a small $<4\times 10^{-4}$ probability.

Our observed velocity of $16.5\pm5.7$ km/s is consistent (in both direction and magnitude) with
the $\sim 10.3$ km/s observed by \cite{sch03}. This again suggests that the
AO and speckle datasets are both detecting real motion of $A_2$ orbiting $A_1$.

\subsection{Is the $\theta^1$ Ori B Group Stable?}

The pair $B_1B_5$ is moving at $\sim 1.4 \pm 4.4$\,km/s in the
plane of the sky w.r.t. to the pair $B_2B_3$ where the escape 
velocity $V_{\mathrm esc} \sim 6$\,km/s for this system. Hence these
pairs are very likely gravitationally bound together. However,
radial velocity measurements will be required to be absolutely
sure that these 2 pairs are bound together.

\subsubsection{Is the Orbit of $\theta^1$ Ori $B_4$ Stable?}

\begin{figure}
\includegraphics[angle=0,width=\columnwidth]{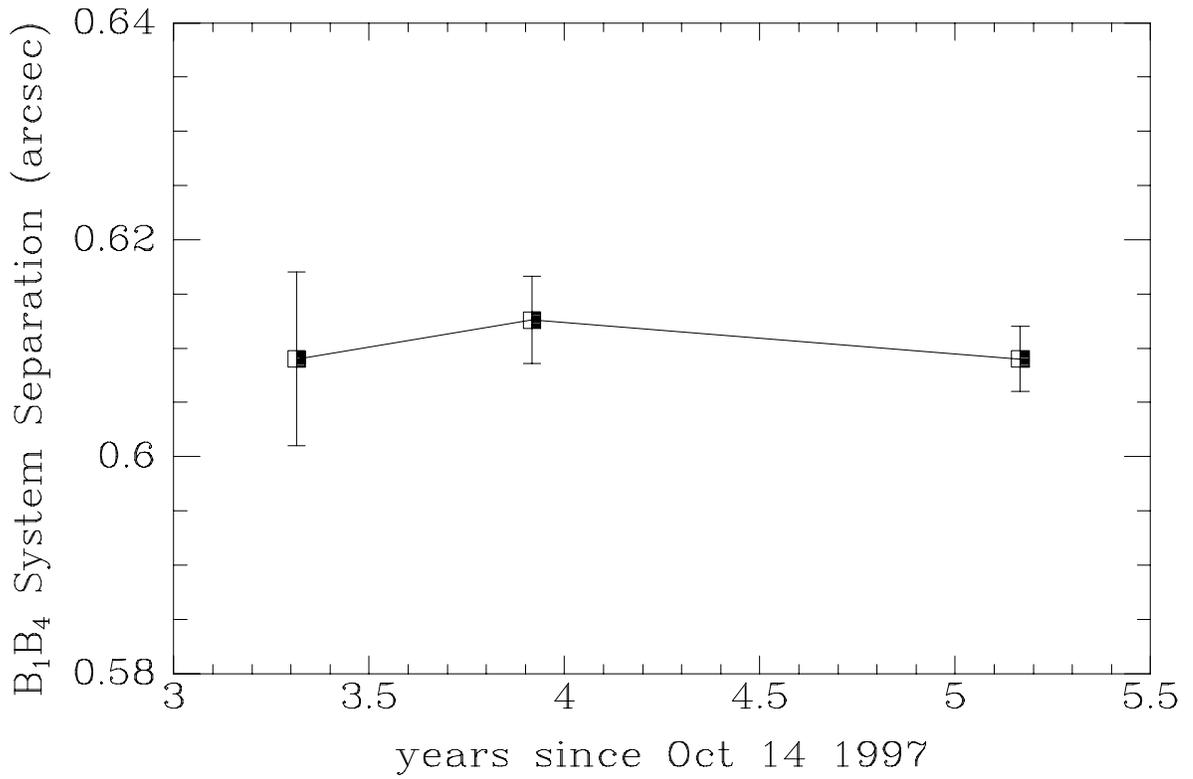}\caption{
The separation between $\theta^{1}$ Ori $B_1$ and $B_4$. Note how over
three years of observation there has been little significant relative
proper motion observed (-0.0017$\pm0.0033\arcsec$/yr; which is
insignificantly different from a constant). If the low mass star $B_4$
is gravitationally bound to the B group the $B_1B_4$ separation should
be roughly constant over these three years. The observed rms scatter
from a constant value is indeed a mere $\pm0.0019\arcsec$, suggesting
the whole $\theta^{1}$ Ori B group is likely physically bound
together. The first data point is an speckle observation from the
6-m SAO telescope
\citep{sch03}, the next point is from our Gemini/Hokupa'a observations
and the last data point is from the MMT AO observations. }
\label{figB4_sep}
\end{figure}

\begin{figure}
\includegraphics[angle=0,width=\columnwidth]{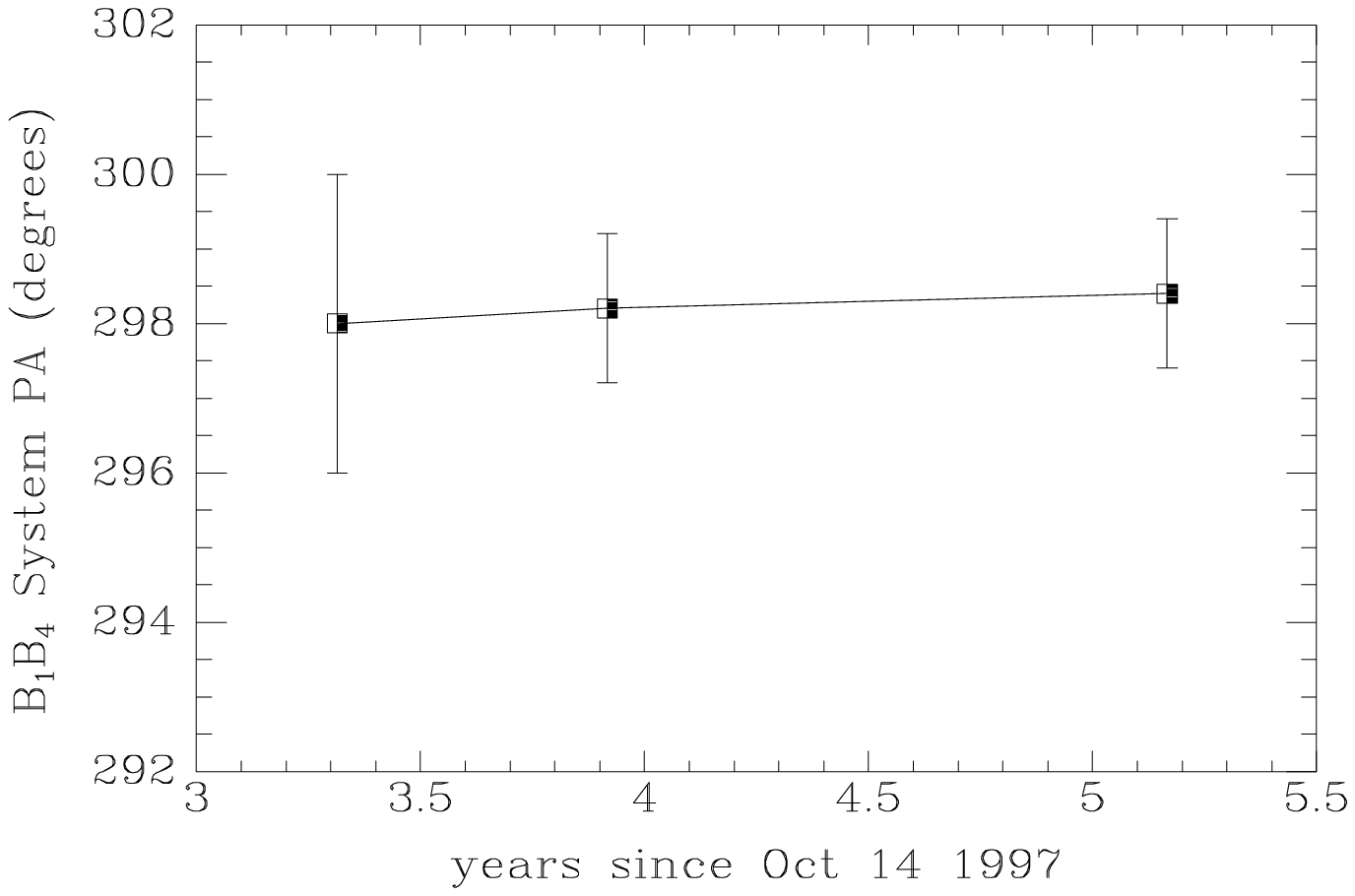}\caption{
The position angle between $\theta^{1}$ Ori $B_1$ and $B_4$. Note how
over three years of observation there has been no significant relative
proper motion observed (0.18$\pm0.9^\circ$/yr which is insignificantly different from a constant). The
error from a constant value is a mere $\pm0.3^\circ$.The first data
point is a speckle observation from the 6-m SAO telescope
\citep{sch03}, the next point is from our Gemini/Hokupa'a observations
and the last data point is from the MMT AO observations. }
\label{figB4_pa}
\end{figure}

The situation is somewhat different for the faintest component of the
group, $B_4$. It has $K = 11.66$ mag which according to Hillenbrand \&
Carpenter (2000) suggests a mass of only $\sim 0.2\,{\mathrm
M}_\odot$. Since there are only 20 stars known to have $K < 11.66$ in
the inner $30 \time 30^{\prime\prime}$ (see Fig.\ 6), we can estimate
that the chances of finding a $K < 11.66$ star within
$0.6^{\prime\prime}$ of $B_1$ is a small $< 8 \times 10^{-3}$
probability. Our two AO measurements (and the one speckle detection of \cite{sch03}) did not detect a
significant velocity of $B_4$ w.r.t. $B_1$ ($4\pm15$ km/s; see Figures \ref{figB4_sep} \& \ref{figB4_pa}). Together with the
escape velocity of $\sim 6$ km/s, this points towards $B_4$ being also
a gravitationally bound member of the $\theta^1$ Ori B group.

On the other hand, its mass and its location w.r.t. to the other four
groups members makes it highly unlikely that $B_4$ is on a stable
orbit within the group. To reconcile these conflicting observations,
one may think of {\it (a)\/} $B_4$'s projected distances from the
other B group members being considerably smaller than the true
distance thus making a stable orbit much more likely, or {\it
(b)\/} $B_4$'s current motion pointing almost exactly along our line
of sight, allowing for a higher true velocity, or {\it (c)\/}
$B_4$ being a chance projection of an object not related to the other
four members of the B group. Without additional astrometric data, we
cannot yet decide which of these three possibilities is the most
likely.

\subsubsection{Is the orbit of $B_3$ around $B_2$ and of $B_5$ around
$B_1$ stable longterm?}

$B_{1}B_{5}$, and $B_{2}B_{3}$ are two binaries with 
projected separations of 0.13 AU ($B_1B_5$) and 52 AU ($B_2B_3$); respectively. The two pairs are separated by a projected distance of 415 AU.
The distance $D_{B_1B_5} \sim 3 \times 10^{-4} \times D_{B_1B_5 B_2B_3}$ and thus
the $B_1B_5$ system is stable. Much more interesting is the case of $B_2B_3$. Their projected distance is not very small compared to their
projected distance (D) from the $B_1B_5$ pair:$ D_{B_2B_3} \sim 0.12 \times D_{B_1B_5 B_2B_3}$. Thus the stability of the $B_2B_3$ orbit needs a more
detailed analysis since it is possible that $B_3$ may be ejected in the future.

\cite{egg95} have given an empirical criterion for the
long-term stability of the orbits of hierarchical triple systems,
based on the results of their extensive model calculations
\citep{kis94,kis94a,egg95}. Their analytic stability criterion is good
to about $\pm 20\%$, and is meant to indicate stability for another
$10^2$ orbits. Given the uncertainties of the masses of the members of
the B group, this accuracy is sufficient for our present discussion.

The orbital period of the two binaries w.r.t. each other is 
$P_{(15)/(23)} \sim 1920\,$yrs, while the orbital period of $B_3$ w.r.t 
$B_2$ amounts to $P_{2/3} \sim 160\,$yrs. For the calculation of both 
periods, we have assumed the masses as given above, and circular orbits 
in the plane of the sky. This leads to a period ratio $X = 
P_{(15)/(23)}/P_{2/3} \sim 12$. Eggelton \& Kiseleva's stability 
criterion requires $X \ge  X_{\mathrm crit} = 10.08$ for the masses 
in the B group. This means that within the accuracy limits of our 
investigation, the binary $B_2B_3$ is just at the limit of 
stability. The stability criterion depends also on the orbits' 
eccentricities. In our case, already mild eccentricities of the order
of $e \sim 0.1$ (as can be expected to develop in hierarchical triple 
systems; see, e.g., Georgakarakos 2002), make the B group unstable.
While we cannot decide yet whether the pair $B_2B_3$ orbit each other
in a stable way, it is safe to say that that the ``triple'' $B_{1}B_{5}$, $B_2$,
and $B_3$ is not a simple, stable hierarchical triple system.

The $\theta^1$ Ori B system seems to be a good example of a highly
dynamic star formation "mini-cluster" which is possibly in the process
of ejecting the lowest-mass member through dynamical decay
\citep{dur01}, and breaking up the gravitational binding of the widest
of the close binaries (the $B_{2}B_{3}$ system). The "ejection" of the lowest-mass member of a
formation "mini-cluster" could play a major role in the formation of
low mass stars and brown dwarfs \citep{rei01a,bat02,dur01,clo03a}. The
breaking up of binaries, of course, modifies the binary fraction of
main sequence stars considerably as well.

\section {Future observations}

In our opinion it is most likely that these $\theta ^1 $ Ori A \& B
group stars are bound. We caution, however, that the motion of each
of these stars could currently be fit equally well by linear motion
(not orbital arcs). Future high resolution observations are required
to see if these stars follow true orbital arcs around each other
proving that they are interacting. In particular, future observations
of the $\theta^1$ Ori $B_4$ positions would help reduce the scatter in
the velocity data and indicate if it is indeed part of the $\theta^1$
Ori B group.

Future observations should also try to determine the radial velocities
of these stars. Once radial velocities are known one can calculate
unambiguously if these systems are bound. Such observations will
require both very high spatial and spectral resolutions. This might be
possible with such future instruments like the future AO fed ARIES
instrument.

\acknowledgements

These MMT observations were possible due to the hard work of the
entire Center for Astronomical Adaptive Optics (CAAO) staff at the
University of Arizona. In particular, we would like to thank Tom
McMahon, Kim Chapman, Doris Tucker, and Sherry Weber for their endless
support of this project. We thank the anonymous referee for helpful
comments that produced a better paper. The Indigo H band filter holder was
installed by graduate student Melanie Freed. The adaptive secondary
mirror is a joint project of University of Arizona and the Italian
National Institute of Astrophysics - Arcetri Observatory. We would
also like thank the whole MMT staff for their excellent support and
flexibility during our commissioning run at the telescope.

The Hokupa'a AO observations were supported by the University of
Hawaii AO group.  (D. Potter, O. Guyon, \& P. Baudoz). Support for
Hokupa'a comes from the National Science Foundation.  These results
were based, in part, on observations obtained at the Gemini
Observatory, which is operated by the Association of Universities for
Research in Astronomy, Inc., under a cooperative agreement with the
NSF on behalf of the Gemini partnership: the National Science
Foundation (United States), the Particle Physics and Astronomy
Research Council (United Kingdom), the National Research Council
(Canada), CONICYT (Chile), the Australian Research Council
(Australia), CNPq (Brazil) and CONICET (Argentina).

The secondary mirror development could not have been possible without
the support of the Air Force Office of Scientific Research under grant
AFOSR F49620-00-1-0294. LMC acknowledges support from NASA Origins
grant NAG5-12086 and NSF SAA grant AST0206351.





\clearpage
\begin{deluxetable}{lllllllll}
\tabletypesize{\scriptsize}
\tablecaption{High Resolution Observations of the $\theta^{1}$ Ori B \& A groups\label{tbl-1}}
\tablewidth{0pt}
\tablehead{
\colhead{System} &
\colhead{$\Delta H$} &
\colhead{$\Delta K^{\prime}$} &
\colhead{Separation} &
\colhead{Separation Vel.} &
\colhead{PA} &
\colhead{PA Velocity} &
\colhead{Telescope} &
\colhead{epoch}\\
\colhead{name} &
\colhead{(mag)} &
\colhead{(mag)} &
\colhead{($\arcsec $)} &
\colhead{(Sep. $\arcsec$/yr)} &
\colhead{($^{\circ}$)} &
\colhead{($^{\circ}$/yr)} &
\colhead{} &
\colhead{(m/d/y)}\\
}
\startdata

$B_{1}B_{2}$&$2.30\pm0.15$&&$0.942\pm0.020\arcsec$&&$254.9\pm1.0$&&SAO\tablenotemark{a}&10/14/97\\
&&$1.31\pm0.10$\tablenotemark{b}&$0.942\pm0.020\arcsec$&&$254.4\pm1.0$&&SAO\tablenotemark{a}&11/03/98\\
&&$2.07\pm0.05$&$0.9388\pm0.0040\arcsec$&&$255.1\pm1.0$&&GEMINI&09/19/01\\
&$2.24\pm0.05$&&$0.9375\pm0.0030\arcsec$&&$255.1\pm1.0$&&MMT&01/20/03\\
&&&&-0.0006$\pm0.0019\arcsec$/yr&&0.07$\pm0.25^\circ$/yr&&\\
\hline
&&&&&&&&\\
$B_{2}B_{3}$&$1.00\pm0.11$&&$0.114\pm0.05\arcsec$&&$204.3\pm4.0$&&SAO\tablenotemark{a}&10/14/97\\
&&$1.24\pm0.20$&$0.117\pm0.005\arcsec$&&$205.7\pm4.0$&&SAO\tablenotemark{a}&11/03/98\\
&&$1.04\pm0.05$&$0.1166\pm0.0040\arcsec$&&$207.8\pm1.0$&&GEMINI&09/19/01\\
&$0.85\pm0.05$&&$0.1182\pm0.0030\arcsec$&&$209.7\pm1.0$&&MMT&01/20/03\\
&&&&$0.0006\pm0.0010\arcsec$/yr &&0.93$\pm0.49^\circ$/yr &&\\
\hline
&&&&&&&&\\
$B_{1}B_{4}$&&$5.05\pm0.8$&$0.609\pm0.008\arcsec$&&$298.0\pm2.0$&&SAO\tablenotemark{c}&02/07/01\\
&&$5.01\pm0.10$&$0.6126\pm0.0040\arcsec$&&$298.2\pm1.0$&&GEMINI&09/19/01\\
&$4.98\pm0.10$&&$0.6090\pm0.0050\arcsec$&&$298.4\pm1.0$&&MMT&01/20/03\\
&&&&$-0.0017\pm0.0033\arcsec$/yr &&0.18$\pm0.95^\circ$/yr &&\\
\hline
&&&&&&&&\\
$A_{1}A_{2}$&$1.51\pm0.15$&$1.38\pm0.10$&$0.208\pm0.030\arcsec$&&$343.5\pm5.0$&&Calar Alto\tablenotemark{d}&11/15/94\\
&&$1.51\pm0.05$&$0.2215\pm0.005\arcsec$&&$353.8\pm2.0$&&SAO\tablenotemark{a}&11/03/98\\
&&$1.62\pm0.05$&$0.2051\pm0.0030\arcsec$&&$356.9\pm1.0$&&GEMINI&09/19/01\\ 
&&&&$-0.0064\pm0.0027\arcsec$/yr &&2.13$\pm0.73^\circ$/yr &&\\
\enddata
\tablenotetext{a}{speckle observations of \cite{wei99}.}
\tablenotetext{b}{these low $\Delta K$ values are possibly due to $\theta^{1}$ Ori $B_{1}$ being in eclipse during the 11/03/98 observations of \cite{wei99}. }
\tablenotetext{c}{speckle observations of \cite{sch03}.}
\tablenotetext{d}{speckle observations of \cite{pet98}.}
\end{deluxetable}





\clearpage

\clearpage

\clearpage

\clearpage

\clearpage

\clearpage

\clearpage

\clearpage

\clearpage

\clearpage

\clearpage

\clearpage

\clearpage

\clearpage

\clearpage

\end{document}